\documentclass[]{aastex631}
\usepackage{graphicx}
\usepackage{txfonts}
\usepackage{verbatim}
\usepackage{color}
\usepackage{natbib}
\submitjournal{ApJ}

\begin{document}

\title{Ultra Long Period Cepheids as standard candles from Gaia to Rubin-LSST.} 

\correspondingauthor{Ilaria Musella}
\email{ilaria.musella@inaf.it}
\author[0000-0001-5909-6615]{Ilaria Musella}
\affiliation{INAF-Osservatorio Astronomico di Capodimonte, Salita Moiariello 16, 80131 Naples, Italy}
\author{S. Leccia}
\affiliation{INAF-Osservatorio Astronomico di Capodimonte, Salita Moiariello 16, 80131 Naples, Italy}
\author{R. Molinaro}
\affiliation{INAF-Osservatorio Astronomico di Capodimonte, Salita Moiariello 16, 80131 Naples, Italy}
\author{M. Marconi}
\affiliation{INAF-Osservatorio Astronomico di Capodimonte, Salita Moiariello 16, 80131 Naples, Italy}
\author{F. Cusano}
\affiliation{INAF-Osservatorio di Astrofisica e Scienza dello Spazio, Via Piero Gobetti, 93/3, I-40129 Bologna, Italy}
\author{M. Di Criscienzo}
\affiliation{INAF – Osservatorio Astronomico di Roma, Via Frascati 33, I-00040 Monte Porzio Catone, Roma, Italy}
\author{G. Fiorentino}
\affiliation{INAF – Osservatorio Astronomico di Roma, Via Frascati 33, I-00040 Monte Porzio Catone, Roma, Italy}
\author{V. Braga}
\affiliation{INAF – Osservatorio Astronomico di Roma, Via Frascati 33, I-00040 Monte Porzio Catone, Roma, Italy}
\author{V. Ripepi}
\affiliation{INAF-Osservatorio Astronomico di Capodimonte, Salita Moiariello 16, 80131 Naples, Italy}
\author{G. De Somma}
\affiliation{INAF-Osservatorio Astronomico di Capodimonte, Salita Moiariello 16, 80131 Naples, Italy}
\affiliation{INAF-Osservatorio Astronomico d’Abruzzo, Via Maggini s.n.c., 64100 Teramo, Italy}
\affiliation{Istituto Nazionale di Fisica Nucleare (INFN) Sez. di Napoli\\
Compl. Univ.di Monte S. Angelo, Edificio G, Via Cinthia \\
I-80126, Napoli, Italy}
\author{M. Gatto}
\affiliation{INAF-Osservatorio Astronomico di Capodimonte, Salita Moiariello 16, 80131 Naples, Italy}
\author{E. Luongo}
\affiliation{Dipartimento di Fisica ”E. Pancini”, Universit\`a degli Studi di Napoli ”Federico II”}
\affiliation{INAF-Osservatorio Astronomico di Capodimonte, Salita Moiariello 16, 80131 Naples, Italy}
\author{T. Sicignano}
\affiliation{Scuola Superiore Meridionale, Largo San Marcellino 10, 80138 Napoli, Italy.}
\affiliation{INAF-Osservatorio Astronomico di Capodimonte, Salita Moiariello 16, 80131 Naples, Italy}
\affiliation{Istituto Nazionale di Fisica Nucleare (INFN) Sez. di Napoli\\
Compl. Univ.di Monte S. Angelo, Edificio G, Via Cinthia \\
I-80126, Napoli, Italy}

\begin{abstract}

An analysis of the Ultra Long Period Cepheids (ULPs) properties could significantly contribute to understanding the Hubble constant tension, e.g. the current discrepancy between determinations based on local distance indicators and those relying on cosmic microwave background measurements. These highly luminous variables are observable beyond 100 Mpc, so if they were confirmed to behave as standard candles, they would allow us a direct measurement of cosmological distances without any secondary distance indicator, thus reducing potential systematic errors in the calibration of the cosmic distance scale.\\
This paper presents an analysis of the largest known sample of 73 ULPs, including 15 objects in nearby galaxies, with new accurate and homogeneous photometry obtained by Gaia DR3, and a new object, in our Galaxy, identified as Long Period Variable in Gaia DR3, but recently reclassified as ULP. The obtained results suggest that, by improving photometric accuracy, the ULP Period-Wesenheit relation shows a smaller dispersion than that obtained in literature and is in better agreement with the Classical Cepheid (CC) one,  supporting the hypothesis that ULPs are the extension of the CCs at higher period, mass and luminosity.
However, to reach this aim, it is necessary to enrich the sample with high-quality data. The Rubin-LSST survey offers the possibility to achieve this thanks to its photometric characteristics and time extension. In particular, we will explore the capabilities of the Rubin-LSST survey to recover ULP theoretical light-curves by using
a new tool called {\tt PulsationStarRecovery}, built by our group for this type of analysis.

\end{abstract}



\section{Introduction}\label{sec:intro}

In recent years, many efforts have been made, both from the theoretical and observational point of view, to understand the origin of the Hubble tension existing between the values obtained for the Hubble constant $H_0$ based on cosmic microwave background (CMB) investigations, coupled with $\Lambda$ Cold Dark Matter theory, and those obtained from the cosmic distance ladder in the local Universe, mainly based on the combination of Classical Cepheids Period-Luminosity ($PL$) relations calibrated geometrically and Supernovae Ia calibrated with CCs. The most recent assessment, based on the latter method, as obtained by A. Riess and his team,  suggests a Hubble constant value of $H_0=74.03\pm 1.04$~km~s$^{-1}$~Mpc$^{-1}$  \citep{Riess+22}. This estimate creates a notable tension, with a $5\sigma$ discrepancy, compared to the Planck CMB value of $H_0=67.4\pm0.5$~km~s$^{-1}$~Mpc$^{-1}$ \citep{Planck+20}.
The CC PL relations allow us to obtain distances out to about 30 Mpc from the space. They are used to calibrate secondary indicators, such as the Type Ia Supernovae, to reach cosmological distances, beyond 100 Mpc. Indeed, at these far distances, the Hubble flow is practically unperturbed and it is possible to measure the Hubble constant. Unfortunately, the use of a ladder with each step calibrated with the previous one causes propagation of the systematic errors in the $H_0$ determination, and, despite numerous attempts to reduce this systematics \citep{yuan+19,reid+19,wong+20,huang+20}, the unresolved $H_0$ tension persists without a clear explanation.
In this context, \citet[][]{Musella21} and \citet{musella22} (hereinafter M21 and M22, respectively) have analysed the properties of all known Ultra Long Period Cepheid (ULPs) variables to assess their reliability as stellar standard candles. Indeed, these pulsators, classified with this name by \citet{Bird+09}, have light curve shapes very similar to the CC ones, of which they have been hypothesised to be the extension at the highest mass and luminosity.  In particular, with their mean luminosity $-9<M_I<-7$~mag, in the Extremely Large Telescopes era \citep{Fiorentino+17}, they should represent the "best" standard candles capable of reaching cosmologically interesting distances, without using any secondary distance indicator, thus reducing the possible effect of systematic errors on the calibration of the extragalactic distance scale and, in turn, on the local determination of $H_0$. On the other hand, as pointed out in previous papers (see M22 and references therein), at these higher masses, the blue loops typical of the CCs are not expected by the evolutionary models and such long period variables are not foreseen by pulsational theories, in particular for very low metallicity values, at the typical blue loop luminosity levels. Therefore, the observational study of these objects is fundamental to improve the capability of evolutionary and pulsational models to understand the nature of these pulsators and their role as standard candles.

The M22 sample is composed by 72 ULPs including all the 18 ULPs of the original sample by \citet{Bird+09} (located within the galaxies LMC, SMC, NGC 55, NGC 300, NGC 6822 and IZw18), 2 objects observed in M81 by \citet{Gerke+11}, 7 ULPs in M31 \citep{Ngeow+15,Taneva+20,Kodric+18}, 2 in M33 \citep{Pellerin+11}, 1 in NGC4151 \citep{yuan+20}, 2 in NGC 6814 \citep{Bentz+19} and 40 ULPs found in the CC samples by \citet{Riess+16} and \citet{Hoffmann+16}, observed in the framework of the SH0ES project \citep{Riess+11} in the galaxies M101, NGC 1015, NGC 1309, NGC 1448, NGC 2442, NGC 3370, NGC 3972, NGC 3982, NGC 4038, NGC4258, NGC 4536, NGC 4639, NGC 5584, NGC 7250, UGC 9391 (hereinafter called SH0ES sample). The SH0ES sample is particularly interesting being photometrically homogeneous. 

To test these pulsators as standard candles, M22 analyzed their reddening independent period-wesenheit ($PW$) relation in the $V$ and $I$ bands \citep[$W_{VI}=I-1.55(V-I)$,][]{Madore82}, in comparison with the CC one. In particular, M22 adopted, as a reference, the CC sample of LMC by OGLE \citep[][]{Sosz+15} and that of NGC 4258 by \citet{Riess+16} and \citet{Hoffmann+16}, the two galaxies assumed as anchors for the extragalactic distance scale. They found that the ULP and CC $PW$ relations are very similar even if that followed by ULPs shows a larger dispersion ($\sigma=0.42$). M22 showed that this spread seems not dependent on the metallicity, but probably intrinsic and/or due to non-accurate photometry, blending effect and the fact that the adopted sample, composed of objects collected in literature, is not photometrically homogeneous. Indeed, adopting only the ULPs in the SH0ES sample \citep{Riess+16,Hoffmann+16}\footnote{ \citet{Riess+16} and \citet{Hoffmann+16} re-analyzed all the Cepheid samples observed in the framework of the SH0ES project, applying a consistent procedure to identify variable stars and their properties.}, the ULP $PW$ relation is in better agreement with the LMC CC one \citep{Sosz+15} and with a reduced dispersion ($\sigma=0.36$).     

From an evolutionary point of view, the $V, I$ color-magnitude diagram of the ULPs shows a broader distribution compared with those of the CCs in LMC and NGC 4258, even when considering only the SH0ES sample, and this behaviour appears to be independent of the metallicity (see M21 and M22 for details).

On this basis, to understand the intrinsic properties of the ULPs and the role of systematic factors, such as reddening corrections and chemical abundance variations, and to get constraints for the evolutionary and pulsational models, we need to have a still more extensive and photometrically homogeneous sample, with multiband photometry extending from the optical to the near-infrared, and covering a broad range of metallicities.

The first possibility to improve M22 analysis comes from the Gaia Data Release 3 (Gaia DR3) Cepheid catalog
\citep{GaiaDR3+23,Ripepi_DR3+23}. This catalog contains new homogeneous photometry for all the known ULPs in the LMC, SMC and M33, for 5 out of 7 ULPs known in M31 (see section \ref{sec:newdata}). In addition, taking advantage of the Gaia DR3 pencil beam survey\footnote{The Gaia DR3 pencil beam survey includes the epoch photometry of all sources (variable and non-variable) in a selected field centered on the Andromeda Galaxy with a radius of 5.5 degree radius.} \citep{Evans_pencil23}, we found the time series for the other two known M31 ULPs included in M22 and identified by \citet{Ngeow+15} and \citet{Taneva+20}. 

Very recently, \citet{Soszynski+24} reported the discovery of the first ULP ($P=78.14$ d) in the Milky Way (MW). This variable has been classified as a Long Period Variable by Gaia DR3 \citep{GaiaTrabucchi+23}, but \citet{Soszynski+24} demonstrated its nature as CC. With this new pulsator, the number of known ULPs increases to 73. Note that the period of this variable is smaller than the lower limit of 80 d fixed for the period of the ULPs by \citet{Bird+09}, but within the typical uncertainties on the period determination. For the ULPs, periods obtained by different authors/data have differences generally within 1-2 d. For example, in the next section, our period re-determination of the MW ULP gives a slightly longer period of 78.42 d. In the same section, we will show the differences among the periods adopted in M22 and those obtained by Gaia DR3 for the variables in common. 
To compare Gaia mean magnitudes with those in the literature, we adopted the transformation from Gaia to Johnson bands obtained by \citet{Pancino+22} also taking into account the correction for the $I$ band, found by \citet{Trentin+23} (for details, see Section \ref{sec:newdata}).

In this context, the forthcoming Rubin-LSST survey \citep[Rubin Observatory Legacy Survey of Space and Time,][]{Ivezic+19} will represent a unique opportunity to improve the photometry and/or increase the sample. The Rubin-LSST survey will cover each region in the southern sky with about 800 images in 6 bands from the ultraviolet to the near-infrared ($ugrizy$) and promises data in the spatial and temporal domains of unprecedented quality. 
The Vera Rubin Observatory involved the scientific community in establishing and refining the survey plan and, in particular, optimizing the observing cadence, testing different scientific cases \citep{Bianco+22}. In particular, our group is investigating the expected results for different types of variables in various Galactic and extragalactic environments with the aim of defining a priority list of the various scientific cases that will benefit from Rubin-LSST data released over time \citep{DiCriscienzo+23,DiCriscienzo+24}.

The organisation of the paper is the following. Section \ref{sec:newdata} describes the new Gaia photometry for the 15 already known ULPs and the new one in the MW. Here, the impact of the new accurate photometry on the ULP PW relation is also discussed in comparison with the CC behaviour. In Section \ref{sec:metric}, we analyse the unique opportunity offered by the Rubin-LSST survey to increase the number of known ULPs and to improve the photometric accuracy of already known ones. To this purpose, we will use a tool developed by \citet{DiCriscienzo+23}, called {\tt PulsationStarRecovery} that, given a LSST cadence strategy, quantifies the precision achieved in the recovery of the light curve and how it varies with the increasing number of visits. Section \ref{sec:conclusions} closes the paper with a summary of the results and some final remarks.    

\section{Gaia data}\label{sec:newdata}

\begin{figure}
\centering
 \includegraphics[width=0.49\linewidth]{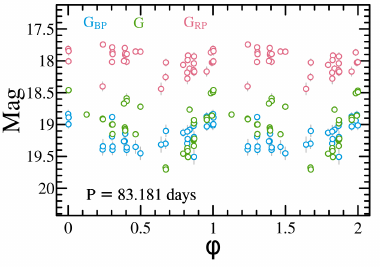}\\
 \includegraphics[width=0.49\linewidth]{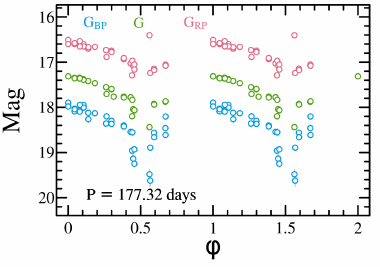}\\
 \includegraphics[width=0.49\linewidth]{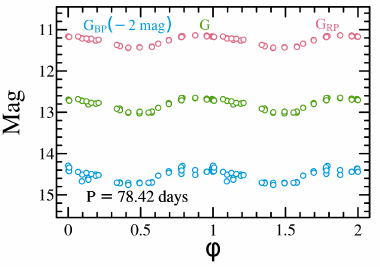}
 \caption{Gaia light curves $G$ (green open circles), $G_{BP}$ (cyan open circles) and $G_{RP}$ (red open circles) for ULPs by \citet{Ngeow+15} (top panel) \citet{Taneva+20} (middle panel) and \citet{Soszynski+24} (bottom panel), respectively. The $G_{BP}$ light curve is shifted of -2 mag. See text for details.}
 \label{fig:lightCurves_Ngeow_Taneva}
\end{figure}

The Gaia DR3 catalog offers the opportunity to have a new homogeneous photometry for the ULPs in the LMC, SMC, and M33 and for 5 M31 ULPs by \citep[][see M22 for details]{Kodric+18}, labelled with ‘‘\_K'' in tables and figures.
For these variables we have the $G$, $G_{BP}$ and $G_{RP}$  intensity averaged mean magnitudes together with new determinations for their period that are slightly different from the previous ones adopted in M22 \citep{Ripepi_DR3+23}.
The time-series for the remaining two M31 ULPs by \citet{Ngeow+15} and \citet{Taneva+20} are included in the Gaia pencil beam survey \citep{Evans_pencil23}. An inspection of their Gaia $G$, $G_{BP}$ and $G_{RP}$ light curves in Fig. \ref{fig:lightCurves_Ngeow_Taneva} shows that the first ULP by \citet[][top panel]{Ngeow+15} is very noisy. For this reason, we derived the Gaia intensity averaged mean magnitudes only for the ULP by \citet{Taneva+20} with a period of 177.32 d (labelled with ‘‘\_T'' in table). In particular, to estimate the mean magnitude and the peak-to-peak amplitude of this variable in the Gaia bands, we fitted the data with a truncated Fourier series computing the intensity average magnitude of the obtained model, as well as the difference between the maximum and the minimum of the obtained model. To estimate the errors in these quantities, we adopted the bootstrap technique. In particular, for each considered band, we simulated a set of 1000 resampled photometric time series repeating the Fourier fit procedure together with the estimate of the mean magnitude and the peak-to-peak amplitude. In the end, the robust standard deviation ($\rm 1.4826\cdot MAD$, where MAD is the median absolute deviation) of the obtained distributions of average magnitudes and amplitudes, is our estimate of the errors on these quantities \citep[see e.g.,][]{Ripepi_DR3+23}. The resulting mean magnitudes and peak-to-peak amplitude with relative errors are reported in the first three lines of Table \ref{tab:ULP_Taneva_MW}. 
For the ULP in the MW, \citet{Soszynski+24}, adopting the OGLE data, found as mean magnitudes $V=16.83$ mag and $I= 11.40$ mag, and report as mean magnitude in the $G$ band the value of 12.84 mag tabulated in the DR3 Gaia catalog. The Gaia light curves for this star are shown in the bottom panel of Fig. \ref{fig:lightCurves_Ngeow_Taneva}. For this pulsator we apply the same procedure described above, obtaining a period of 78.42 d, slightly different from that published by \citet{GaiaTrabucchi+23} and adopted by \citet{Soszynski+24}, and the mean magnitudes and amplitudes in the $G_{BP}$, $G$ and $G_{RP}$ reported in the last three rows of the table \ref{tab:ULP_Taneva_MW} with the relative errors. Our determination for the mean $G$ band is 0.031 mag smaller than that by \citet{GaiaTrabucchi+23}. The distance modulus of this new ULP can be determined from its Gaia parallax $0.176 \pm 0.077$ mas. The error on the parallaxes is about 40\% and in these cases, the distance can be determined by adopting a probabilistic approach \citep[and references therein]{GaiaBayler24}. Adopting this method, \citet{GaiaBayler24} found for this variable a photo-geometric distance (that takes into account also the magnitude of the object) of $3829^{-330}_{+304}$ pc corresponding to an absolute distance modulus of $12.91^{-0.19}_{+0.17}$ mag.

\begin{table}
    \centering
    \begin{tabular}{ccc}
    \hline
       Gaia band   & mean magnitude & amplitude\\
       & mag& mag \\
       \hline
       \multicolumn{3}{c}{M31 ULP by \citet{Taneva+20}}\\
       \hline
       $G_{BP}$  & $18.379 \pm 0.050$   & $0.905\pm0.068$  \\
       $G$      & $17.652 \pm 0.020$ & $0.690\pm0.080$\\
       $G_{RP}$  & $16.824 \pm 0.045$   & $0.493 \pm 0.076$\\
       \hline
       \multicolumn{3}{c}{MW ULP by \citet{Soszynski+24}}\\
       \hline
       $G_{BP}$  &$ 16.539 \pm 0.010$ &  $0.360 \pm 0.023$\\
       $G$  &$12.809 \pm 0.004$ &  $0.379 \pm 0.013$\\
       $G_{RP}$  & $11.271 \pm 0.003$  &$0.312 \pm 0.009$\\
   \hline
    \end{tabular}
    \caption{Gaia mean magnitudes and peak-to-peak amplitudes for the ULP by \citet{Taneva+20} and \citet{Soszynski+24}.}
    \label{tab:ULP_Taneva_MW}
\end{table}

\begin{figure}
    \centering
    \includegraphics[width=0.49\linewidth]{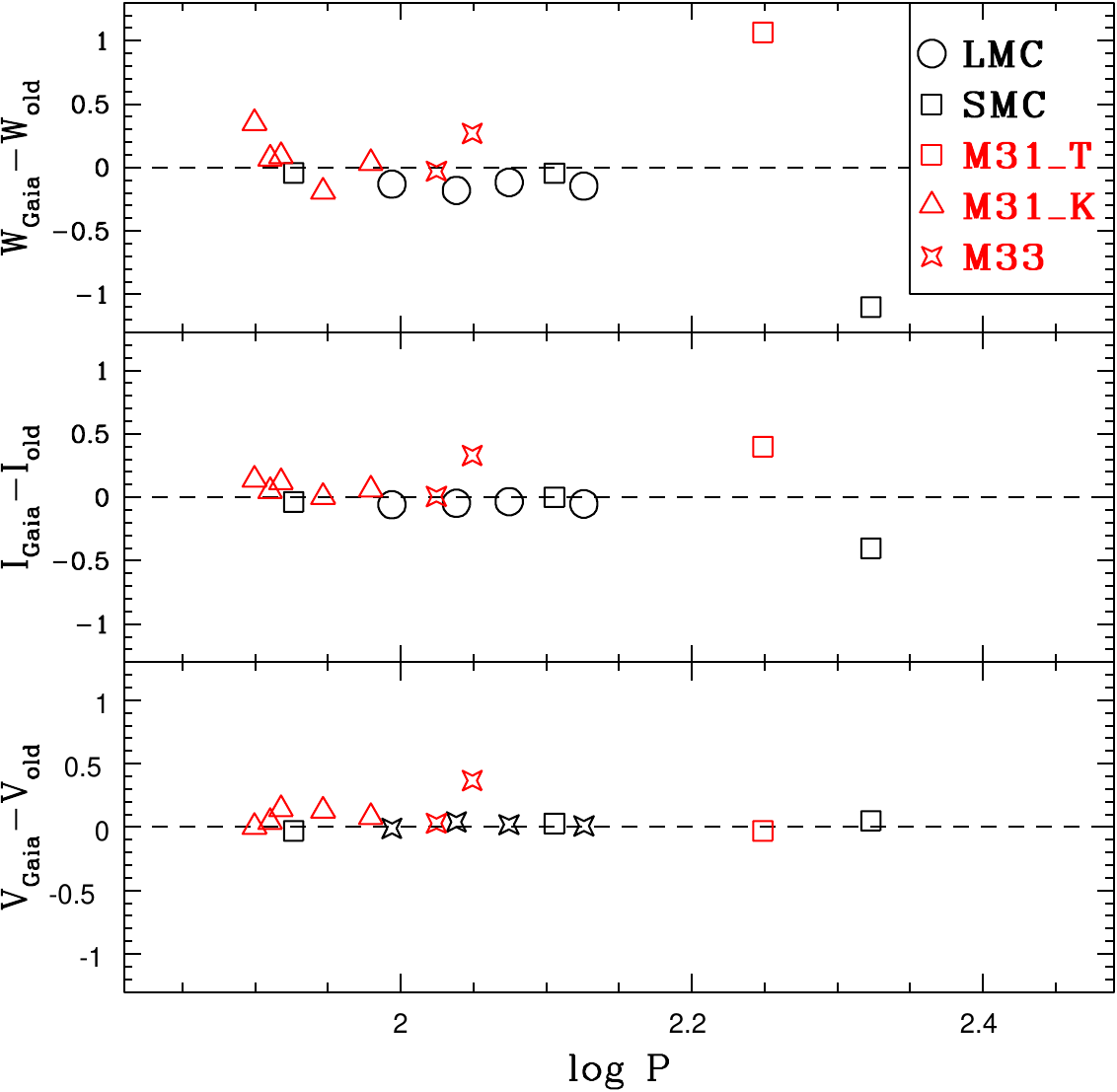}
    \caption{Differences between $V_{Gaia}$ and $I_{Gaia}$ mean magnitudes obtained in this paper and the ones adopted in M22 (labelled as $old$), for the ULPs included in Table \ref{tab:newdata}. See text for details.}
    \label{fig:diff_gaia}
\end{figure}

\begin{figure}
\begin{minipage}[c]{0.7\textwidth} 
\centering
    \includegraphics[width=\textwidth]{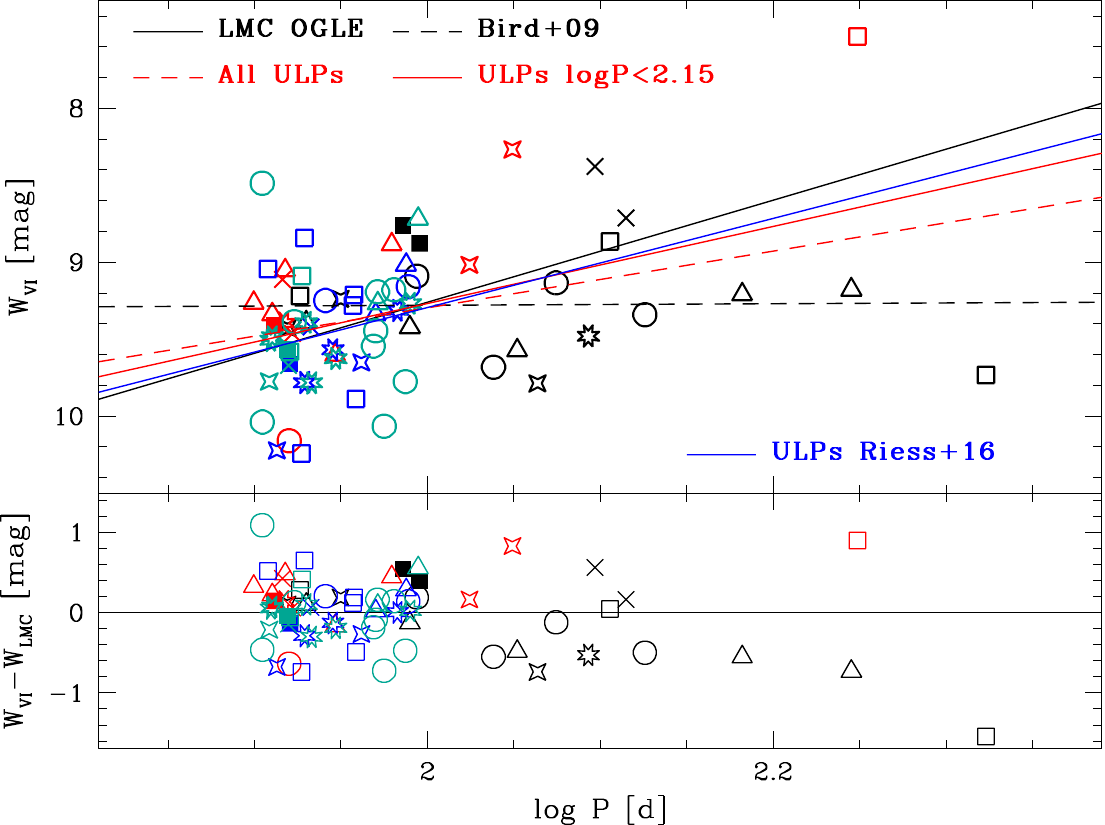}\\
    \includegraphics[width=\textwidth]{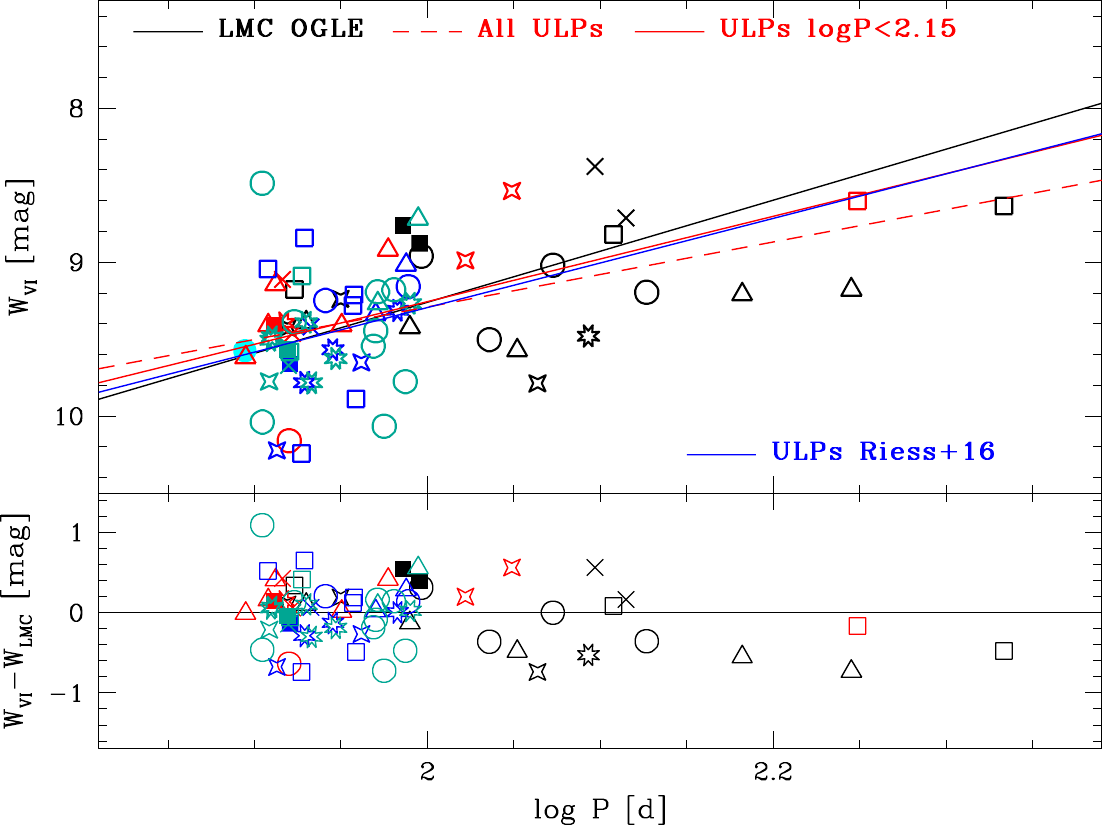}  
\end{minipage}
\hspace{5mm}
\begin{minipage}[c]{0.2\textwidth}
\centering
\includegraphics[width=0.95\textwidth]{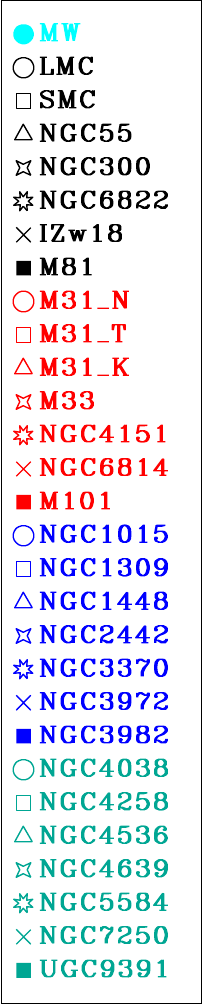}
\end{minipage}
    \caption{Comparison between the ULP $PW$ relations obtained by M22 (top panel) and those derived in this paper (bottom panel) for the different selections adopted by M22 (red solid line for all ULPs, red dashed line including only ULPs with $\log{P}<2.15$ and blue solid line for the homogeneous ULP sample by \citealt{Riess+16}) and compared to the OGLE LMC CC one (black solid line) and the \citet{Bird+09} ULP relation (black dashed line). Note that the position of the new MW variable cyan filled dot in the bottom panel) is in perfect agreement with the CC and ULP relations. See text for details.}
    \label{fig:comparW}
\end{figure}

The Gaia mean magnitudes were transformed into the Johnson-Cousins $V$ and $I$ magnitudes adopting the transformations by \cite{Pancino+22}, including the additional correction of 0.035 mag found by \citet{Trentin+23} (see their fig. 1). The obtained $V_{Gaia}$ and $(V-I)_{Gaia}$ values are reported in Table \ref{tab:newdata}. For the variables already studied in M22, in Table \ref{tab:newdata} we also report the new periods \citep{Ripepi_DR3+23} and the previous values for $P$, $V$, and $(V-I)$ (labelled with $old$) adopted in M22. The differences obtained for the $V$, $I$ magnitudes and the relative reddening independent $W$ function are plotted in Fig. \ref{fig:diff_gaia}. We notice that in particular for the longest period ULPs, the differences for the $W$ magnitudes amount to more than 1 magnitude. This result significantly impacts the $PW$ relation obtained including all the known ULPs. In Fig. \ref{fig:comparW}, we compare the old $PW$ figure by M22 (top panel) with the one based on the updated photometry (bottom panel). In particular, in both figures, we have, in the upper plot, the $W_{VI}$ of the ULPs placed at the LMC distance and, in the bottom plot, the differences between the observational $W_{VI}$ of the ULPs and that obtained by adopting the OGLE relation \citep[ $W_{LMC}$,][]{Sosz+15}, plotted as a function of the $\log{P}$.

The resulting $PW$ relations for the different selections adopted by M22 are plotted and compared with the OGLE LMC CC one and the \citet{Bird+09} ULP relation. The effect of relying on more accurate and homogeneous magnitudes is well visible in the definition of the $PW$ relation, especially at longer periods, where the dispersion is reduced. 

It is worth noting that the values $V_{Gaia}$ and $I_{Gaia}$ derived for the MW ULP are different from those obtained by \citet{Soszynski+24} but confirm a color $(V-I)$ much larger than for the other ULPs (4.832 mag). As already pointed out by \citet{Soszynski+24}, this ULP is located near the MW plane and close to the Galactic bulge, where a very high extinction is expected (they found $A_{Ks}\thickapprox 0.84$ mag). On the other hand, \citet{Soszynski+24} discussed the possibility that this variable was a different type of pulsator, analyzing all the possible alternatives, but confirming its nature as CC. An additional constraint is obtained in this paper, given that the position of this variable in the reddening independent $PW$ diagram is in perfect agreement with the CC and ULP relations.

The new $PW$ relation obtained including all the ULPs is $W_{VI} = -2.11 (\pm 0.49) \log P - 4.98 (\pm 0.96)$ with $\sigma=0.36$ mag, a smaller dispersion than that obtained by M22 ($\sigma=0.42$ mag). Despite this improvement, as already pointed out in M22, the period range $\log P>2.15$ is poorly sampled, probably due to the challenging strategy required to identify and characterize variables with such very long periods. As these missing points can critically affect the determination of the slope, a more solid result is expected including only the ULPs with $\log P<2.15$ for which we find $W_{VI} = -2.77 (\pm 0.76) \log P - 3.69 (\pm 1.49)$ with $\sigma=0.36$, in better agreement with the CC one \citep[with a slope of $-3.314 \pm 0.008$ and $\sigma=0.077$ mag, based on a sample of 2455 objects,][]{Sosz+15} than that obtained in M22. In addition, we find a very good agreement with the slope of $-2.89 \pm 2.05$ and $\sigma=0.36$ obtained by M21 adopting only the photometrically homogeneous sample by \citet{Riess+16}. The results obtained from this updated analysis seem to indicate that improving the photometric accuracy and homogeneity of the sample implies a better agreement of the ULP properties with the CC ones and a reduced dispersion of the relation, supporting the hypothesis that these variables are the counterpart of CCs at higher mass, luminosity and period.

\begin{table}
    \centering
    \begin{tabular}{lrrcccc}
    \hline
Galaxy & $P\ \ \ $   & $P_{old} $ &   $V_{Gaia}$ &   $(V-I)_{Gaia}$  &  $V_{old}$   &   $(V-I)_{old}$ \\ 
& d$\ \ \ $ & d$\ \ \ $ &mag &mag&mag&mag\\
\hline
MW &  78.42 & ---& 16.311  &  4.832 & --- & ---\\
LMC & 118.2  &   118.70  &    12.01 &    1.173     &   11.99       &      1.120           \\
LMC & 108.6  &   109.20  &    12.45 &   1.156      &   12.41       &      1.070  \\         
LMC & 99.2  &    98.60  &    11.91 &   1.157      &   11.92       &  1.110              \\  
LMC & 133.9  &   133.60  &    12.13 &   1.151      &   12.12       &  1.090                \\
SMC & 215.5  &   210.40  &   12.33  &   1.281      &     12.28     &      0.830            \\
SMC & 128.1  &   127.50  &    11.95 &   1.059      &   11.92       &  1.030              \\  
SMC & 83.8  &    84.40  &    11.94 &   0.915     &    11.97      &     0.910             \\
M31\_T& 177.32 &  177.32  &   18.13  &    1.400      &  18.16        &   1.830               \\
M31\_K& 78.5 &  79.35 &   18.79 &    1.260 &    18.79  &  1.398 \\
M31\_K& 80.9 &  81.35   &   19.12  &   1.471      & 19.08         & 1.483              \\  
M31\_K&  81.7&  82.74   &   19.21  &    1.610      & 19.07         & 1.591                \\
M31\_K&  89.2&  88.45   &   19.14  &   1.478      & 19.01         & 1.352                \\
M31\_K&  94.9&  95.38  &   19.21  &   1.698      & 19.13         & 1.681                \\
M33 &   105.2&  105.80 &  18.29 &    1.191     & 18.25       &   1.168  \\
M33 &   111.9 &  111.98 &  18.19 &   1.333      & 17.83        &  1.294  \\
\hline
    \end{tabular}
    \caption{$V_{Gaia}$ and $(V-I)_{Gaia}$ values obtained 
    from the Gaia mean magnitudes, adopting the transformations by \cite{Pancino+22}, including the additional correction of 0.035 mag found by \citet{Trentin+23}. For the variables already studied in M22, we also report the previous values (labelled with $old$). See text for details.}
    \label{tab:newdata}
\end{table}

 On this basis, in the next section, we explore the results expected from the Rubin-LSST survey for the ULPs to obtain a sample statistically significant and photometrically homogeneous.

\section{Light curve's recovery of Local Group ULPs with Rubin-LSST}\label{sec:metric}

Rubin Observatory involved the scientific community in a process to set and refine the Rubin-LSST observing strategy through the analysis of different scientific cases \citep{Bianco+22} and our group is contributing analyzing the capability of Rubin-LSST for the study of various types of pulsating stars in different environments 
\citep{DiCriscienzo+23,DiCriscienzo+24}. In this context, this work is focused on the possibility to improve and/or increase the sample of known ULPs through this survey.

In particular, we used a tool called {\tt PulsationStarRecovery} \citep{DiCriscienzo+23} that provides a simulation of the Rubin-LSST time series starting from a template and that quantifies the accuracy of the recovery of the light curve's period and shape (mean magnitude and amplitude) as a function of the different simulated observational strategies and the number of years of the survey.

In detail,  {\tt PulsationStarRecovery}  uses a pulsating star template curve (in our case, a theoretical light curve; see below) as input. This template is positioned at specified sky coordinates, defined by Right Ascension and Declination, with a given distance modulus and reddening. Based on this information, the tool generates simulated photometric time series in multiple bands ($ugrizy$). These simulations are based on the sampling and signal-to-noise ratio at the specified sky position, as provided by the chosen Rubin-LSST survey strategy simulation, while accounting for the distance modulus and reddening. Subsequently, {\tt PulsationStarRecovery} analyzes the simulated photometric time-series to identify the pulsation period using a multi-filter approach \citep[\texttt{Gatspy},][]{Gatspy} and computes the best-fit model for the simulated light curves in each photometric band. The tool provides several outputs that evaluate the quality of light curve recovery from the comparison of the results (period, amplitude, mean magnitude) with those of the input theoretical model \citep[for more details, see][]{DiCriscienzo+23}.

\begin{figure}
    \centering
    \includegraphics[width=0.9\linewidth]{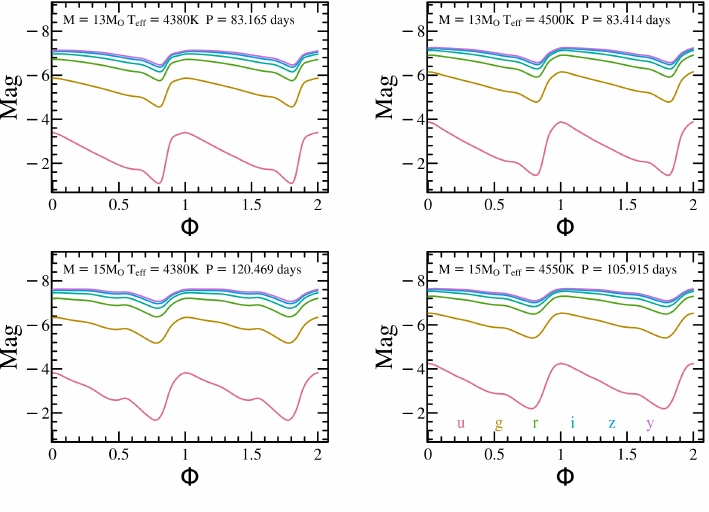}
    \caption{Light curves in the Rubin filters of the theoretical pulsating star models in Table \ref{tab:TheoLC}.}\label{fig:TheoLC}
\end{figure}

In the following, we adopt the Rubin-LSST survey strategy simulation baseline\_v3.0\_10yrs.db, describing the most probable observing strategy as outlined in the Phase 2 document (\url{https://pstn-055.lsst.io/}).
As input for the {\tt PulsationStarRecovery}, we adopt four theoretical light curves characterized by different stellar parameters (mass, effective temperature, and luminosity) expected by the ULPs \citep[][M21 and M22]{Fiorentino+12,Fiorentino+13}, with a period spanning from 80 to 120 days. These theoretical light curves have been computed by adopting non-linear, convective pulsation models \citep[see for example][]{Bono+99,Marconi+17}, whose main characteristic is the possibility of predicting all the pulsation observables, namely periods, instability strip boundaries, light, radius, radial velocity and temperature curves, mean magnitudes and colors, pulsation amplitudes. The theoretical bolometric light curves can be transformed into the Rubin-LSST filters ($ugrizy$) using stellar bolometric corrections, such as the ones provided by \citet{Chen+19}. Table \ref{tab:TheoLC} reports the stellar properties of the adopted models (identified with the label in column 1) together with the mean magnitudes and peak-to-peak amplitudes in the Rubin-LSST bands, and Fig. \ref{fig:TheoLC} shows the relative light curves in these filters.

\begin{table}
   \centering        
   \footnotesize
     \begin{tabular}{ccccccccccccccccc}
    \hline
     Input   & $M$ &$T_{eff}$  & $\log{L/L_{\odot}}$ & $P$ & mean $u$ & amp $u$ & mean $g$ & amp $g$& mean $r$ & amp $r$& mean $i$ & amp $i$& mean $z$ & amp $z$& mean $y$ & amp $y$\\
      Model & $M_{\odot}$ & K & & d & mag & mag & mag & mag& mag & mag& mag & mag& mag & mag& mag & mag\\
    \hline
     A &   13 &4380  & 4.46 & 83.165 &$-$2.56 & 2.29&	$-$5.42 &	1.31 &$-$6.41	&0.96 &	$-$6.72&	0.82&	$-$6.87&	0.72&	$-$6.95&	0.66\\
     B &    13 &4500  & 4.5 & 83.414 & $-$2.92	& 2.41&	$-$5.62&	1.36&	$-$6.55	&0.99	&$-$6.84&	0.85	&$-$6.98	&0.73	&$-$7.04	&0.68\\
     C &    15 &4550  & 4.67 & 105.915& $-$3.03 &	2.14 & $-$5.93&	1.17&	$-$6.94&	0.84&	$-$7.25&	0.71 	&$-$7.40	&0.60	&$-$7.47	&0.55\\
     D &    15 &4380  & 4.67 & 120.469 & $-$3.41&	2.08 &	$-$6.08 &	1.14 &	$-$7.00 &	0.82 &	$-$7.28 &	0.70 &	$-$7.41 &	0.60 &	$-$7.47 &	0.56\\
     \hline
    \end{tabular}
    \caption{Stellar properties of the adopted theoretical pulsating stars models and mean magnitudes and peak-to-peak amplitudes in the Rubin-LSST filters. See text for details.}
    \label{tab:TheoLC}
\end{table}

As shown above, Gaia allows us to have more accurate data for the ULPs in the LMC, SMC and M31.
To test the capability of the Rubin-LSST survey to extend these results to more distant Local Group galaxies, we have analized the recovery of ULP light curves, starting from LSST time series, in the only three galaxies hosting known ULPs and observable from Vera Rubin Observatory NGC 6822, NGC 300 and NGC 55. Unfortunately, the other galaxies that host ULPs and are more distant than NGC 55 and visible from Cerro Pachòn fall all below Rubin's detection limit (the SHOES sample's nearest galaxy NGC 4536 is located at $\mu=30.91$ mag). On the other hand, ULPs in MCs have magnitudes beyond the saturation threshold. 

The properties of the selected galaxies in which we analized the recovery of ULP  light curves starting from LSST time series, namely their equatorial coordinates, the adopted distance modulus and color excess, and the number of known ULPs taken from M22 are summarized in Table \ref{tab:selgal}.

\begin{table}
    \centering
    \begin{tabular}{cccccc}
    \hline
         Galaxy & RA& DEC & $mu_0$ & $E(B-V)$ & N. ULPs\\
                &  deg & deg& mag   &  mag    & \\    
    \hline            
         NGC 6822    & 296.240592 & -14.803434 & 23.31  &     0.21  &  1 \\    
         NGC 300 & 13.723 & -37.684 & 26.37 &  0.01&  3\\
         NGC 55  & 3.723 & -39.197& 26.43 &  0.01& 5 \\
    \hline     
    \end{tabular}
\caption{Local Group galaxies selected for ULP recovery with {\tt PulsationStarRecovery}.}
    \label{tab:selgal}
\end{table}

\begin{figure}
    \centering
    \includegraphics[width=0.46\linewidth]{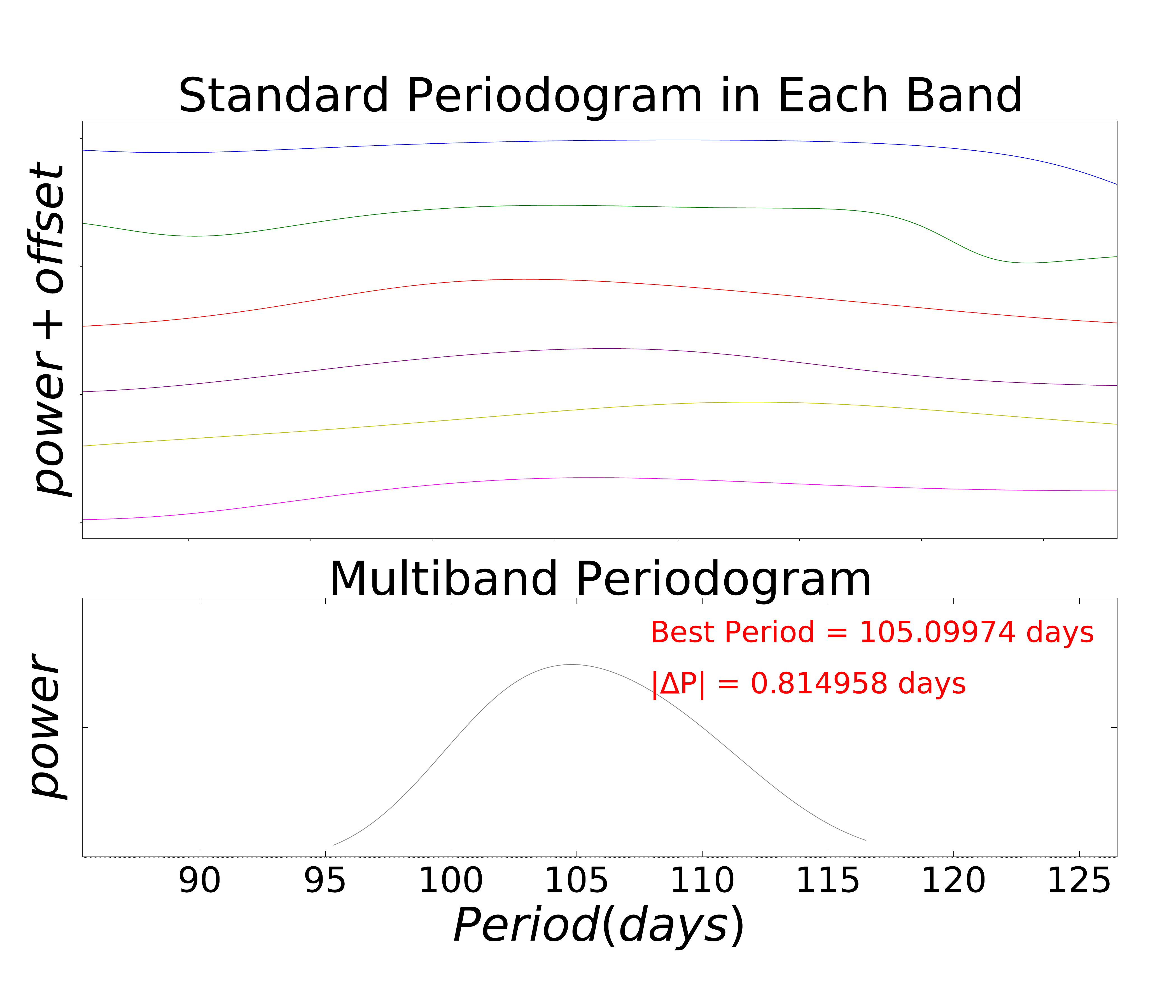}\hfil
    \includegraphics[width=0.46\linewidth]{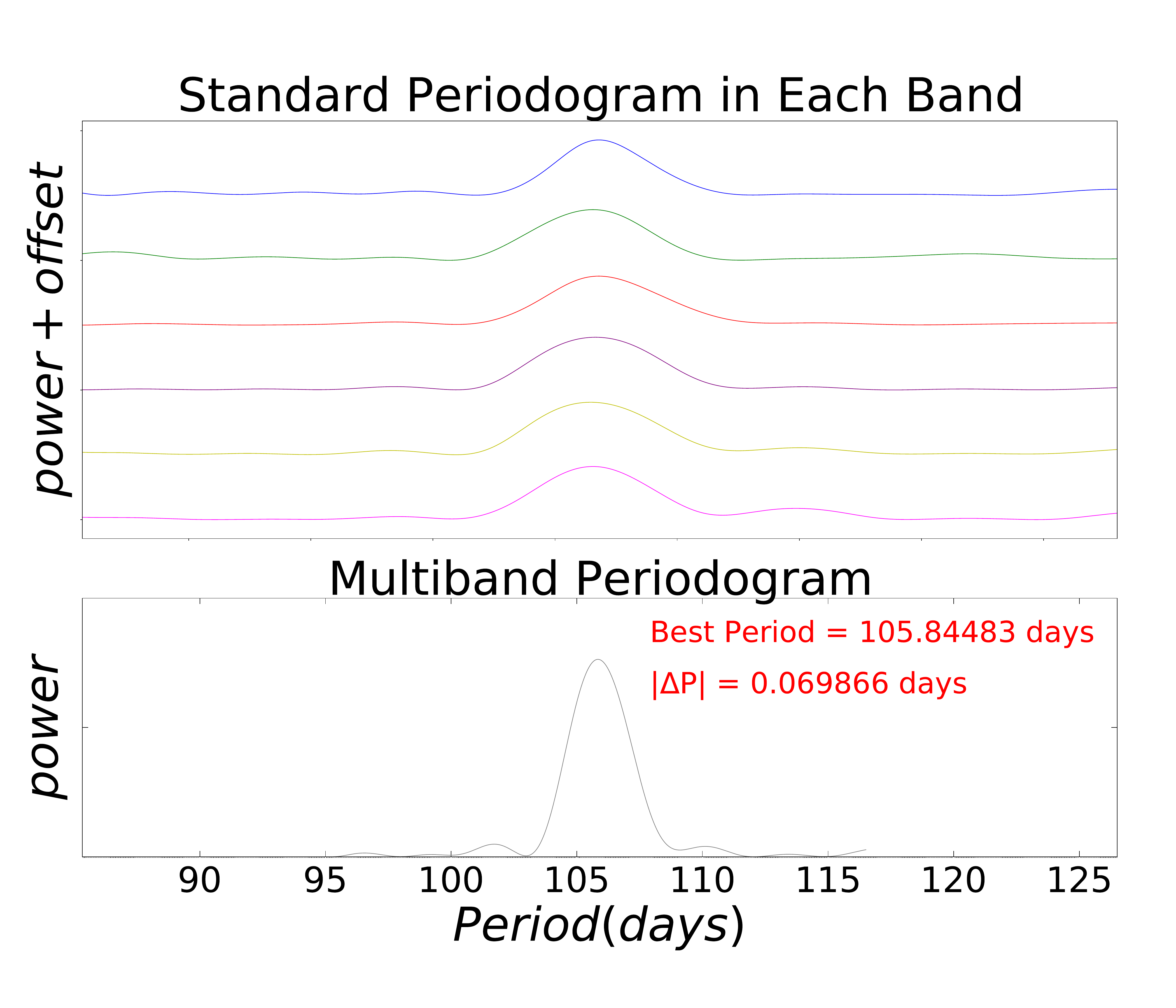}\\
    \vspace{0.5cm}
    \includegraphics[width=0.49\linewidth]{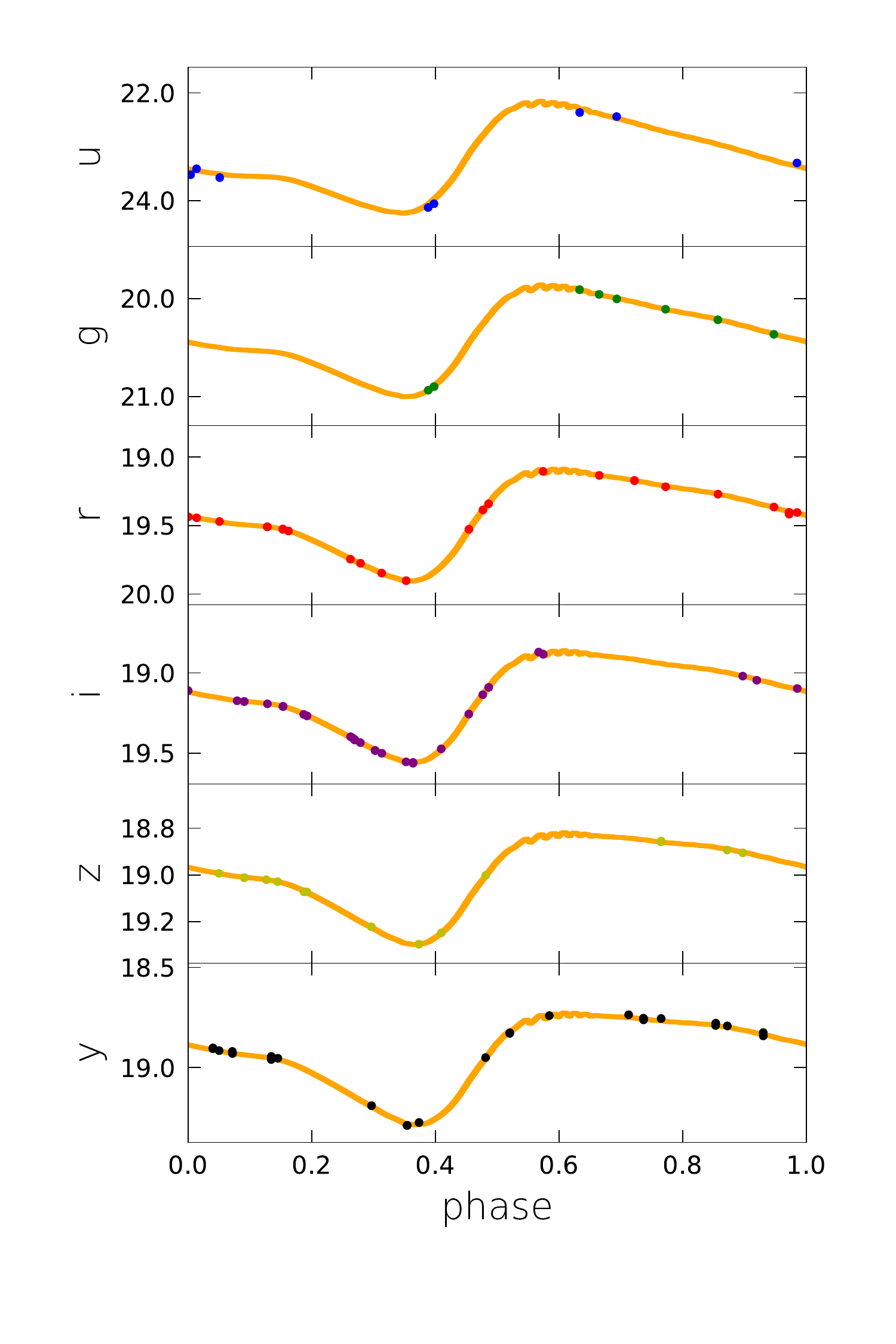}\hfil
    \includegraphics[width=0.49\linewidth]{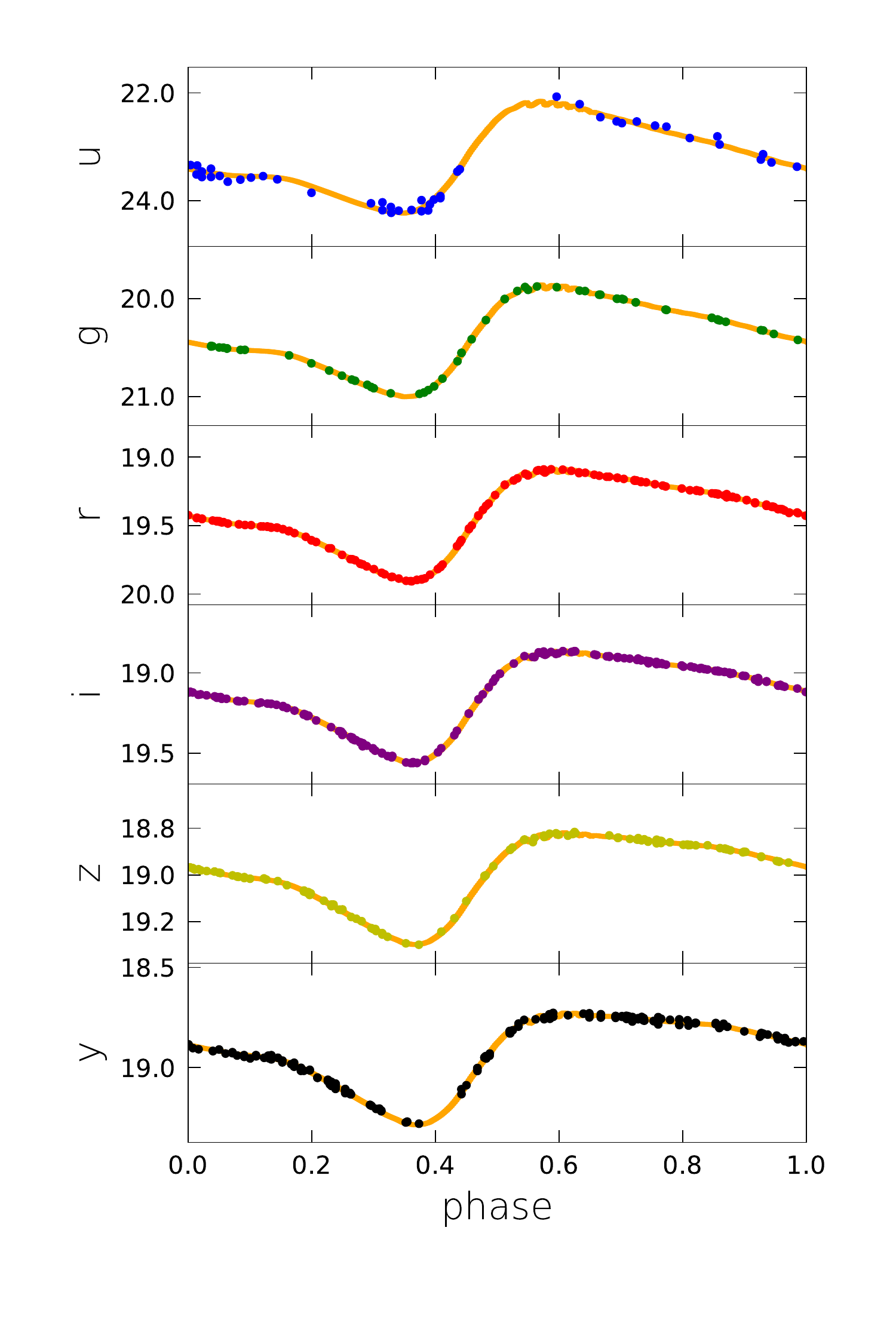}
    \caption{Recovered period and light curve  as a function of the years for model C in NGC300.  The top panels show the periodograms obtained after 2 (on the left) and 10 (on the right) years of the survey, respectively, using the single bands separately (colored solid lines) or adopting a multi-filter approach (black solid line). The bottom panels display the simulated phase points and the input models (orange line) in all the bands after 2 (on the left) and 10 (on the right) years, respectively, obtained by adopting the corresponding periods. See text for details.}
    \label{fig:metric}
\end{figure}

\begin{figure}
    \centering
   
   \includegraphics[width=\linewidth]{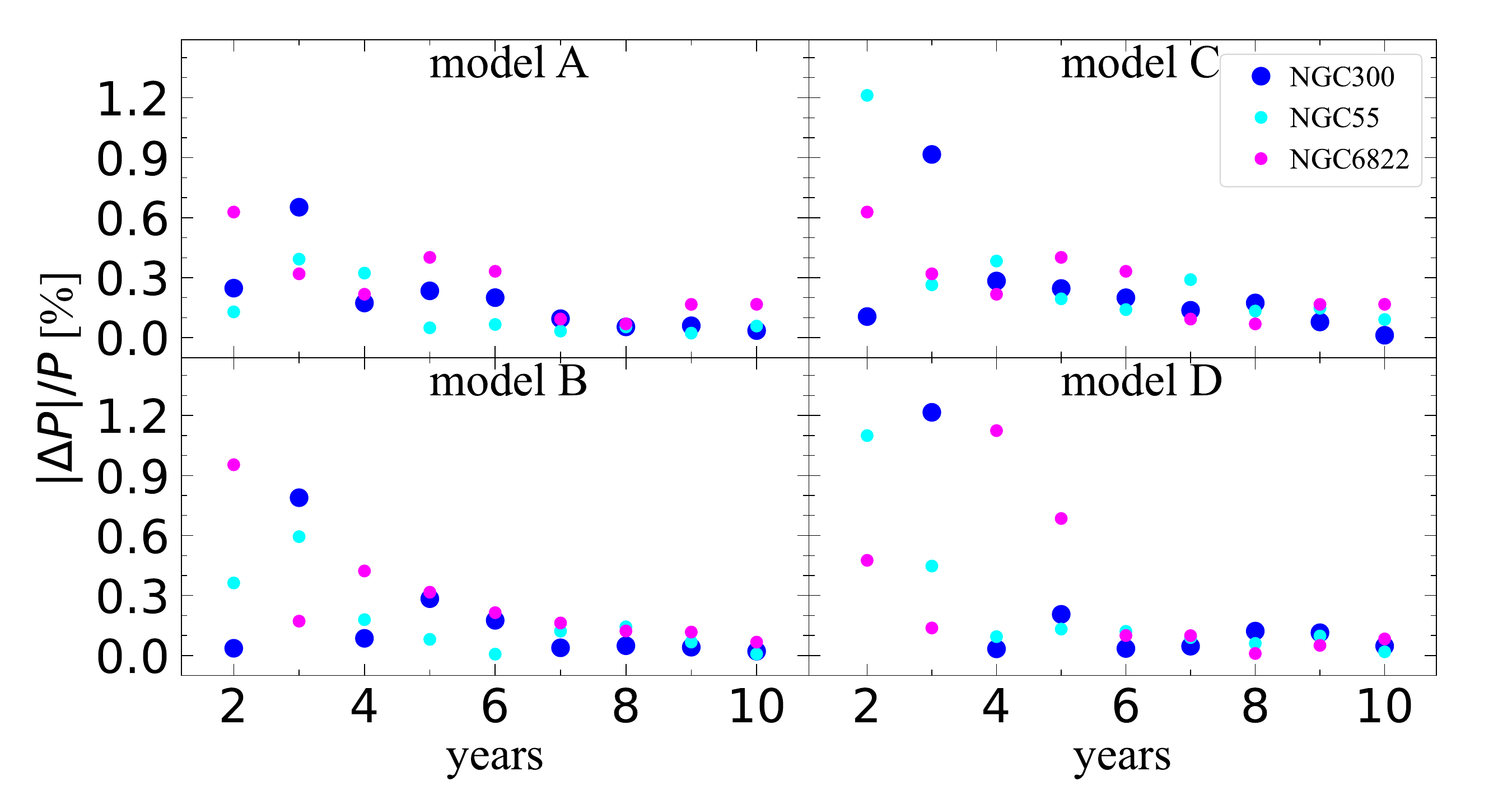}
   
    \caption{Period recovery. See text for details.}
    \label{fig:NewdeltaP}
\end{figure}

\begin{figure}
    \centering
    \includegraphics[width=\linewidth]{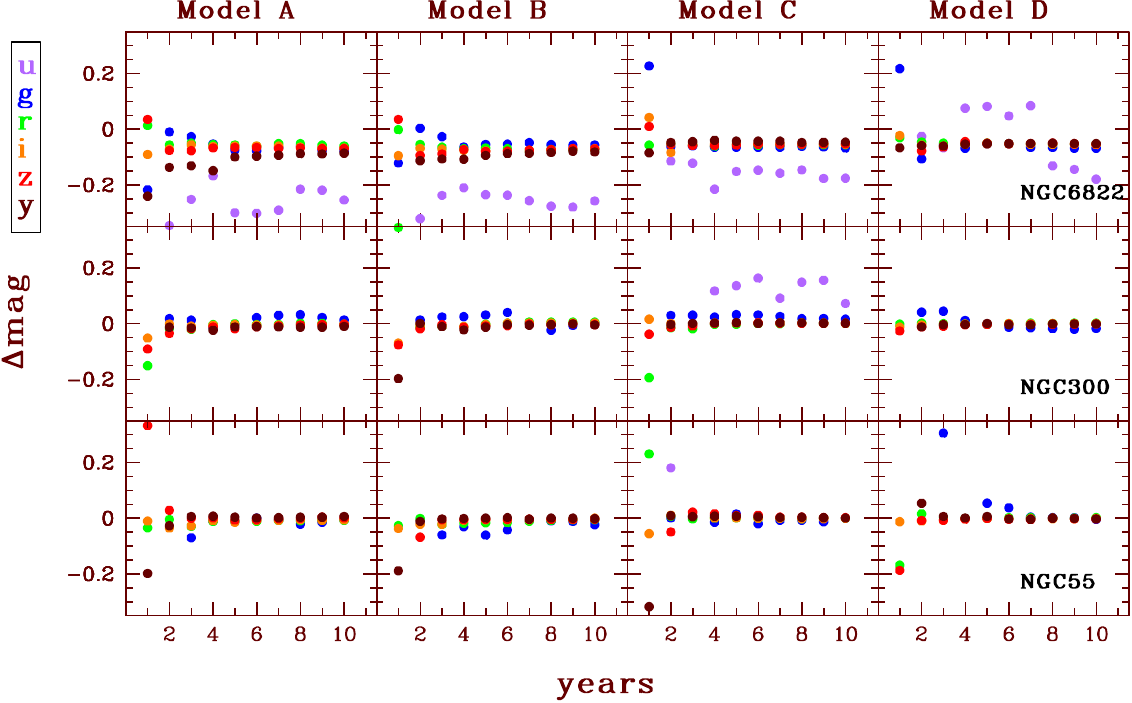}
    \caption{Mean magnitude recovery for $ugrizy$ bands. See text for details}
    \label{fig:Dmag}
\end{figure}

\begin{figure}
    \centering
    \includegraphics[width=\linewidth]{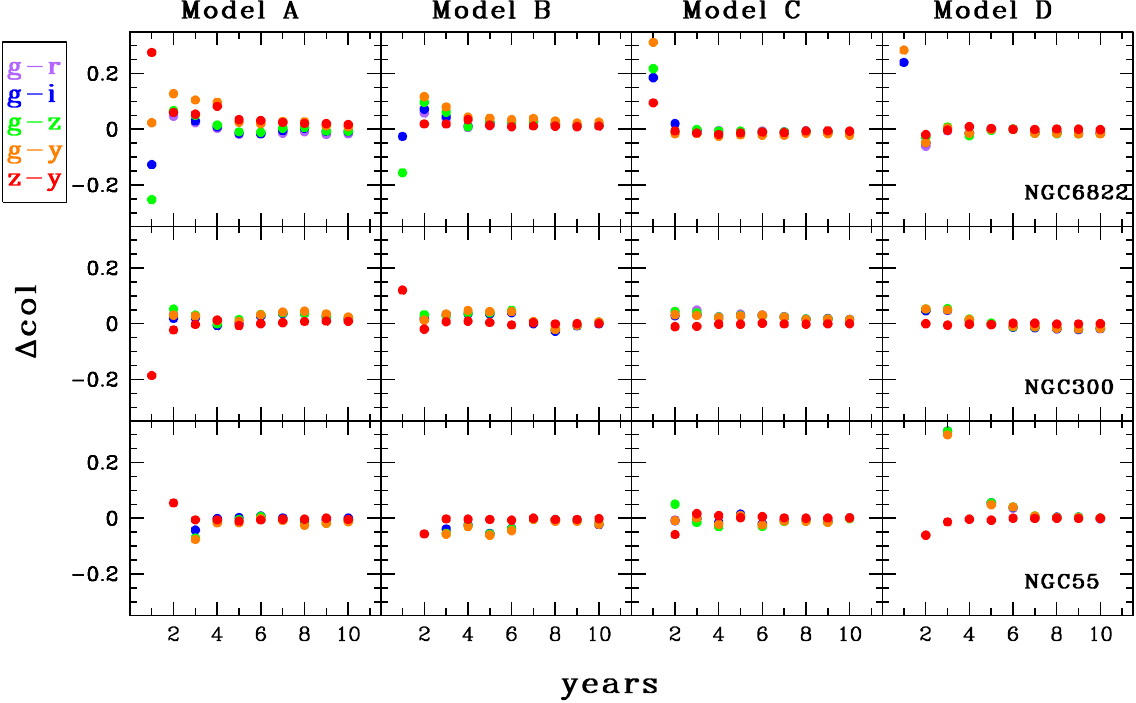}
    \caption{Color recovery for $g-r$, $g-i$, $g-z$, $g-y$ and $z-y$. See text for details.}
    \label{fig:Dcol1}
\end{figure}

\begin{figure}
    \centering
    \includegraphics[width=\linewidth]{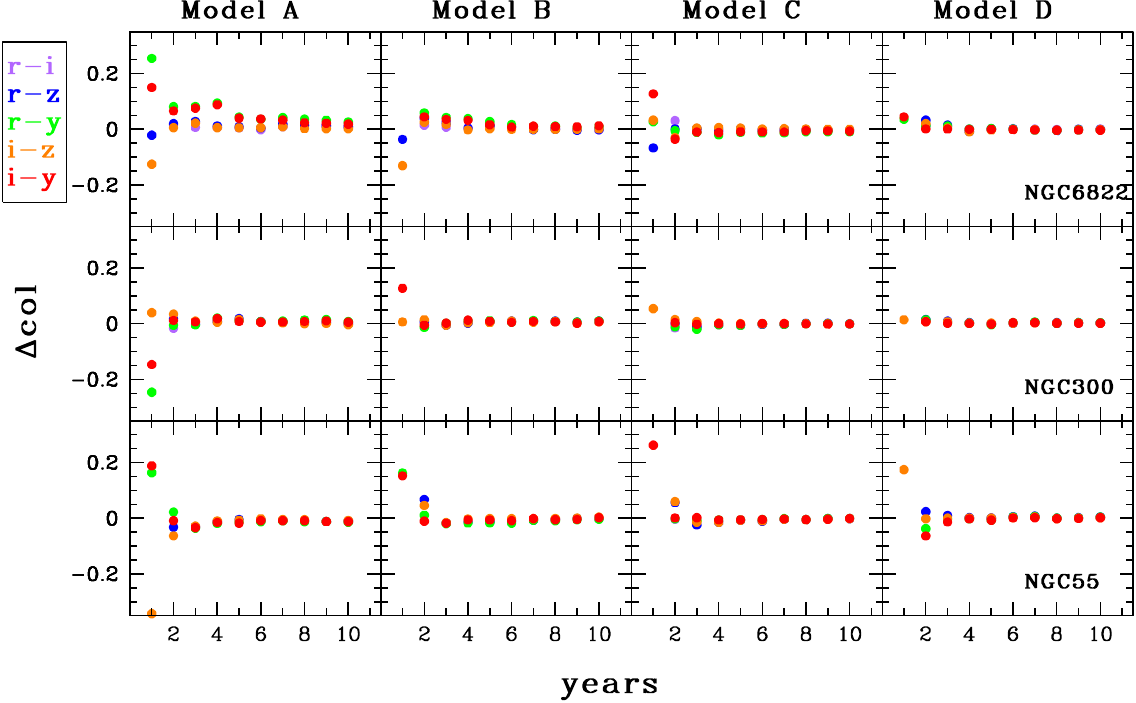}
    \caption{Color recovery for $r-i$, $r-z$, $r-y$, $i-z$ and $i-y$. See text for details.}
    \label{fig:Dcol2}
\end{figure}

\begin{figure}
    \centering
    \includegraphics[width=\linewidth]{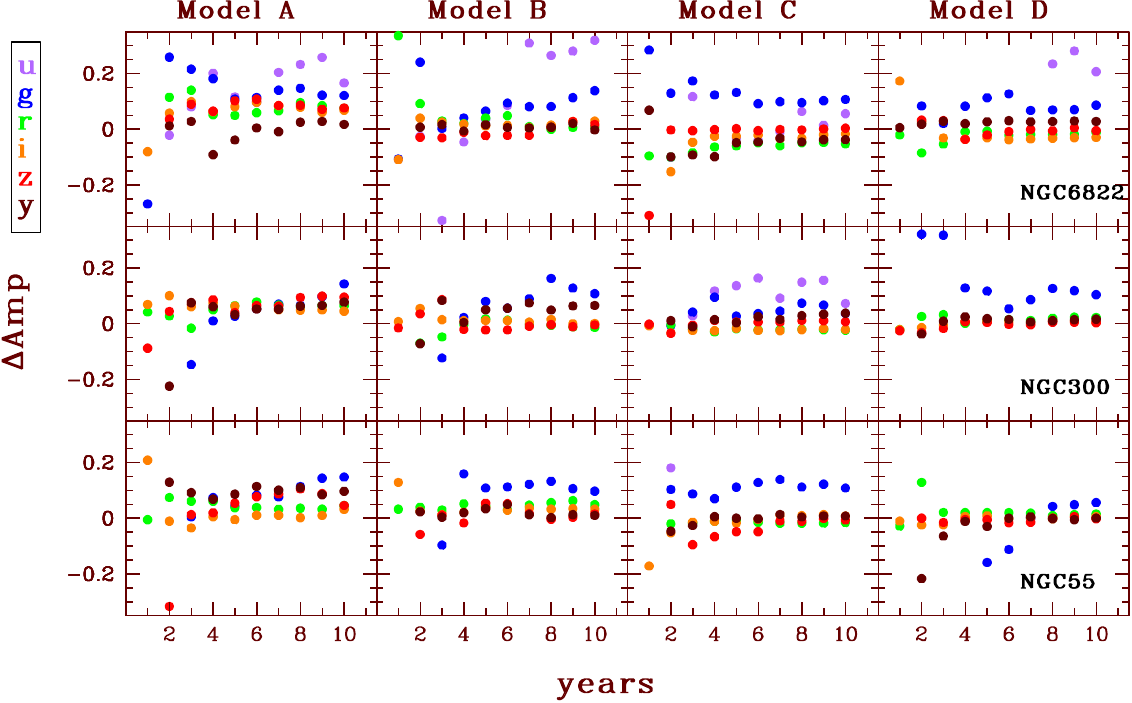}
    \caption{Amplitude recovery for $ugrizy$ bands. See text for details}
    \label{fig:Damp}
\end{figure}

As an example of how the tool {\tt PulsationStarRecovery}  works, we consider in input the ULP light curve model C (see Table \ref{tab:TheoLC}) assumed to be located in the galaxy NGC 300. Fig. \ref{fig:metric} shows the results of the simulations, adopting the Rubin-LSST survey strategy baseline\_v3.0\_10yrs.db. The top panels show the periodograms obtained after 2 and 10 years of the survey, using the single bands separately or adopting a multi-filter approach. The bottom panels display the simulated phase points and the input models in all the bands after 2 and 10 years, obtained by adopting the corresponding periods from the previous step. These multi-filter light-curves are fitted adopting a truncated Fourier series of cosines, computing the best-fit model and deriving the peak-to-peak amplitudes and the zero points of the fits (representing the mean magnitude) in all the bands. In addition, {\tt PulsationStarRecovery} produces a complete dictionary including several outputs to quantify the quality of the recovery of the input light curve \citep[for more details see][]{DiCriscienzo+23}. The results of our analysis for 4 different input theoretical ULP models in 3  different galaxies are summarized in Figs. \ref{fig:NewdeltaP}, \ref{fig:Dmag}, \ref{fig:Dcol1}, \ref{fig:Dcol2}, and \ref{fig:Damp}.

Figure \ref{fig:NewdeltaP} shows the dependence of the period recovery ($\vert \Delta P\vert/P$ in percentage) on the number of years of the survey for the various models. In this plot, we exclude the first year of the survey which is not enough for a sound recovery of the input period, but, starting from the second year, the result is reliable and improves as the number of years increases. It is worth noting that the dependence on the sky coordinates and distance modulus is negligible at least in the distance range considered.

Figures \ref{fig:Dmag}, \ref{fig:Dcol1}, \ref{fig:Dcol2} and \ref{fig:Damp} show the differences between theoretical mean magnitudes, colors and peak-to-peak amplitudes of the models adopted as input and the recovered values (see Table \ref{tab:TheoLC}). The panel columns correspond to the different models (with the ULP period increasing from left to right) and the panel rows correspond to the different galaxies (with distance modulus increasing from top to down). The results obtained for the various bands/colors are plotted in different colors and are shown only in the range from -0.4 to 0.4 mag, excluding larger differences. 
The first evidence is that, for all the cases explored, the $u$ band is not reliable to have an accurate determination of the mean magnitude and peak-to peak amplitude because the number of visits is never sufficient to reconstruct the light curve. For the other bands, the results show that the average magnitude is recovered with an accuracy of about $0.1$ mag already from the second year. Also concerning the colors, we achieved very good results after 2-3 years of the survey, for all the adopted band combinations.

Regarding the peak-to-peak amplitude, also the $g$ filter presents some problems, probably due to the large peak-to-peak amplitudes in this band and due to the lower number of visits compared to the other bands. The results for the other bands give an error within 0.2 mag for the nearest galaxies decreasing to 0.1 mag only after 4 years. This occurrence is due to the difficulty of reconstructing the light curve shapes with only a few points \citep{DiCriscienzo+23}.

It is important to note that concerning the use of ULPs in the cosmic distance ladder and to constrain their behaviour in the Color-Magnitude diagrams, we essentially need only mean magnitudes, colors and periods, even if, for studies related to their pulsational and evolutionary properties, we have to take into account the uncertainties found in the amplitude recovery.  Beyond the information that will be obtained at the end of the survey, which is optimal for characterizing the typical variability of ULPs and long period variables in general, the key result to highlight here is that Rubin data will be able to recover ULP light curves in the Local Group with sufficient precision to conduct impactful science starting from the very first releases. This is a very crucial but quite unexpected result.

To complete this analysis, we need to account for the crowding/blending effects, due to the possible association of CCs and ULPs to crowded and high-surface-brightness regions. 
The crowding is not included in the simulation of the Rubin-LSST survey that considers stars as isolated objects \citep{Yoachim+16}. This effect depends on the adopted wavelength and becomes more and more important when increasing the distance of the studied galaxy (due to decreased angular resolution)  \citep[see also the discussion in][]{Fiorentino+13a}.


In a recent paper \citep{DiCriscienzo+24}, our group analyzes the effect of the crowding on the RR Lyrae in the Galactic Bulge, using the same Rubin-LSST simulations adopted in this paper, but including the TRILEGAL simulation for the star distribution in our Galaxy. The results show that, if the blending is not recognized as such by the photometry software, the effect is to measure a brighter luminosity and a smaller amplitude, with the consequence of inferring  shorter distance. On the contrary the crowding has no appreciable effects on the determination of the period.

On this basis, we would need to have an estimate of the error due to crowding. Unfortunately, while this evaluation is relatively easy in our galaxy \citep{DiCriscienzo+24}, it is certainly more challenging in external galaxies, where accurate simulations of the radial profile of the stellar population would be needed. This effect obviously varies from galaxy to galaxy. \citet{Riess+23} pointed out that to accurately quantify the mean local surface brightness, we need to use artificial star tests.

To roughly estimate the blending effect on the ULPs, we produced the simulated light curves also including a possible blending  considering an additional flux to the model flux: $flux_{blendend}=flux_{model}+X \times <flux_{model}>$,  where X is a percentage and  $<flux_{model}>$ is the mean value of the $flux_{blendend}$. This is a simplistic treatment that is valid in the hypothesis that the blending objects have the same colour as the analyzed variable, but it can be acceptable to be able to understand the capabilities of our tool to recover the ULP light curves in crowded fields.
Considering, for example, the results obtained by \citet{Moche+01} for the blending of Cepheids in M33, we expect, in particular for variables with periods longer than 10 days, an average blending of 10-20\% in flux, depending on the considered filter, even if, for example, \citet{Bresolin+05} pointed out that in a sample of 16 Cepheids in NGC 300, only three are significantly affected by blending. 
In any case, in our test, we considered the input model C, in the galaxy NGC 55, the farthest galaxy in our sample, and adopted $X=10, 30\%$ and, as an extreme case, $X=50\%$. Considering the previous results, we exclude the $u$ band from this analysis because the total number of visits expected in this band is too few for a good light curve recovery. 

\begin{figure}
    \centering
    \includegraphics[width=0.5\linewidth]{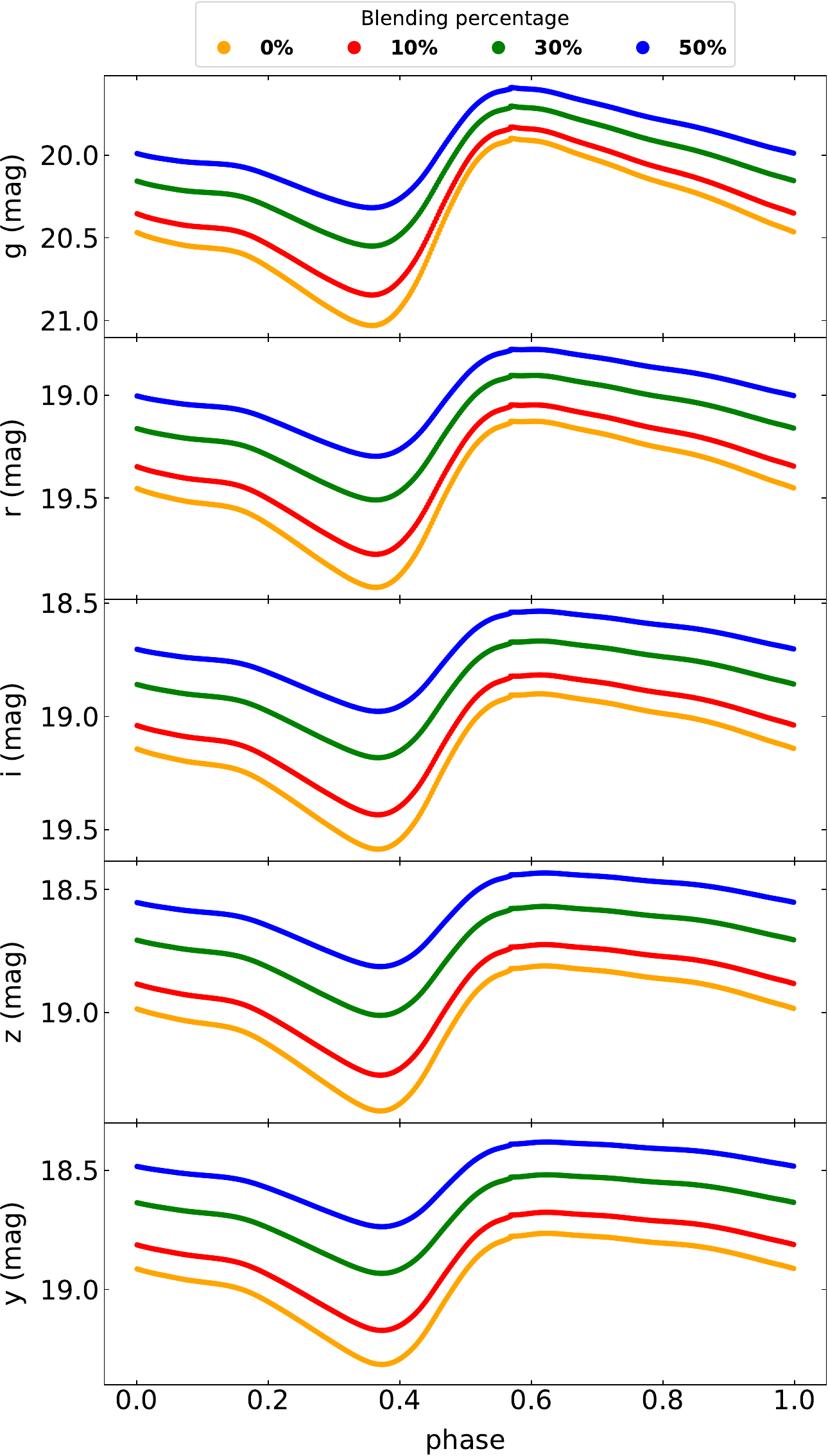}
    \caption{The effect of a blending of 10\%, 30\% and 50\% on the recovered light curves for the model C (see Table \ref{tab:TheoLC}) located in the galaxy NGC 55. See text for details}
    \label{fig:LCblend}
\end{figure}

\begin{figure}
    \centering
    \includegraphics[scale=1]{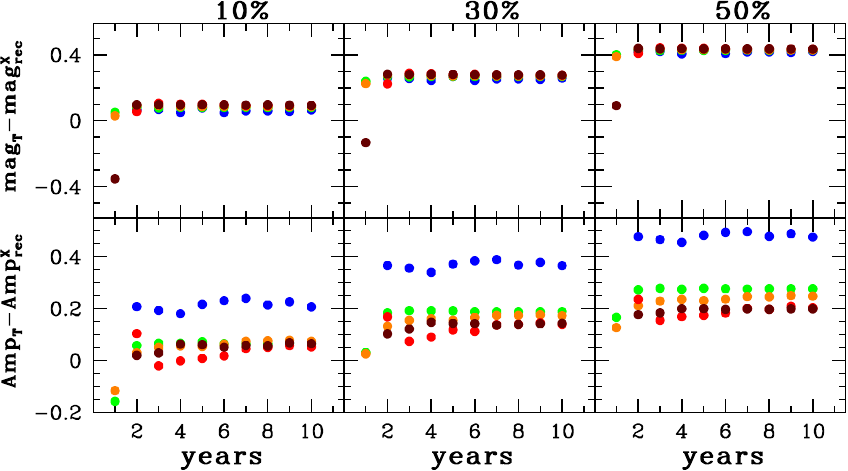}
    \caption{The effect of a blending of 10\% (left panels), 30\% (center panels) and 50\% (right panels) on the recovered magnitude (upper panels) and amplitudes (bottom panels) for the model C (see Table \ref{tab:TheoLC}) located in the galaxy NGC 55. The differences plotted correspond to the model value ($mag_{T}$ and $Amp_{T}$) minus the recovered one for each considered blend ($mag_{rec}^{X}$ and $Amp_{rec}^{X}$). Color code is the same in Fig. \ref{fig:Damp}.}
    \label{fig:blend}
\end{figure}

The effect of the blending on the model light curve is reported in Fig. \ref{fig:LCblend} and the results for mean magnitudes and amplitudes are summarized in Fig. \ref{fig:blend}.  The differences plotted correspond to the model values without any blend ($mag_{T}$ and $Amp_{T}$), minus the recovered ones for each considered blend ($mag^{X}_{rec}$ and $Amp^{X}_{rec}$). As expected, when the blending factor increases, we have a shift in the mean magnitudes towards smaller magnitudes and a reduction of the amplitude. In any case, also in the $50\%$ extreme case, recovered amplitudes are still large enough to allow us to identify new ULPs. Concerning the mean magnitudes, our experiment indicates that the errors on the recovered values (and thus on the distances) due to the blending can be significant, indicating the necessity of estimating the crowding on the real data. This estimate could be obtained through artificial star test experiments that are very time-consuming for a large survey like Rubin-LSST. For this reason, as already pointed out by \citet{DiCriscienzo+24} and \citet{DalTio+22}, these results suggest the need to have in the LSST science photometric pipelines an efficient package capable of properly deblending stars in crowded regions \footnote{\url{https://pipelines.lsst.io/v/d_2024_08_27/modules/lsst.meas.extensions.scarlet/overview.html}} and to add a measurement of the errors due to crowding to LSST data products adopting approximate methods \citep[see e.g.][]{Olsen+03}.

In any case, on this basis, thanks to the capability of the Rubin-LSST survey to recover the ULP light curves in the Local Group, we expect to increase the number of known objects and improve the statistics and the accuracy of the data for the already known ULPs, in a few years.

\section{Conclusions}\label{sec:conclusions}

This paper is part of a project aiming to analyze the possibility of defining the ULPs as standard candles capable of reaching the Hubble flow in a single step. In particular, previous papers by \citet{Bird+09}, \citet{Fiorentino+13}, M21 and M22 pointed out the hypothesis that these pulsators represent the counterpart of the CC at higher luminosity and mass. M21 and M22 concluded that a larger sample of ULPs with more accurate and homogeneous photometry was required to reduce residual errors and uncertainties and to get constraints to improve pulsational and evolutionary models. In this paper, we relied on the Gaia DR3 catalogue \citep{GaiaDR3+23, Ripepi_DR3+23} and the Gaia pencil beam survey \citep[on the galaxy M31,][]{Evans_pencil23} to improve the photometry of 15 already known ULPs in the Local Group galaxies LMC, SMC, M31 and M33 and to get the photometry of an object belonging to the Milky Way, recently classified as ULP by \citet{Soszynski+24}. Our results demonstrate that an improvement of the photometry accuracy reduces the errors on the ULP $PW$ relation and improves the agreement with the CC one, further supporting the hypothesis that they are the same type of pulsating variables but with different mass and period ranges.

In addition, we discussed the unique opportunity offered by the Rubin-LSST survey to enlarge the population of identified ULPs in the Local Group with accurate and homogeneous photometry. This work is part of a more general analysis carried on to contribute to the optimization of the observing cadence of the Rubin-LSST survey, testing different scientific cases \citep{Bianco+22}. In particular, our group is involved in the investigation regarding several types of pulsating stars in different environments \citep{DiCriscienzo+23,DiCriscienzo+24}. Indeed, thanks to the tool {\tt PulsationStarRecovery} \citep{DiCriscienzo+23}, we can anticipate the potential outcomes of the Rubin-LSST survey and optimize the scientific exploitation of the forthcoming data. This tool, taking into account the Rubin-LSST observational strategy,  measures the goodness of the recovery of the pulsational properties (period, amplitudes, mean magnitudes, and colors) as a function of the number of years of the survey. 
In particular, we have adopted 4 different theoretical models for ULPs characterized by different mass, luminosity, and period and analyzed the recovery of their pulsational properties in the 3 Local group galaxies hosting known ULPs observables by Rubin-LSST, namely NGC 6822, NGC 300 and  NGC 55. 

From our analysis, we have found that:
\begin{itemize}
    \item we can obtain a very good recovery (errors smaller than 1-2\%) of the input period starting from the second year of the survey with negligible dependence on the period, sky coordinates and distance;
    \item the $u$ band is not reliable to derive accurate mean magnitudes and amplitudes;
    \item without considering the uncertainties due to the crowding, the mean magnitudes in the $grizy$ bands are recovered with an error $\le 0.1$ mag from the second year. To this error, we should add the uncertainty due to the crowding/blending. 
    \item the recovery of the amplitudes in the different bands has larger uncertainties, in particular for the $u$ and $g$ bands and in the first years of the survey. The recovery worsens in the presence of blending.
    \item A rough estimation of the crowding effect shows that notwithstanding the reduction of the amplitudes due to the blending, Rubin-LSST will be able to identify new ULPs in the Local Group, in particular in the $gri$ bands. However, the effect on the recovery of the average magnitude can become significant, highlighting the importance of including
crowding-related errors as data products in each release of Rubin-LSST.
    
\end{itemize}

On this basis, the expected results from the Rubin-LSST survey, to extend to the Local Group the results obtained from the Gaia survey in the Magellanic Clouds and in M31 and M33, are encouraging. Above all, this work shows that it will not be necessary to wait until the end of the survey to address the open questions about ULPs, indeed already from the first releases we will have data useful to better understand their use as standard candles, both from a theoretical and observational point of view. Finally, it should be noted the importance of a tool like {\tt PulsationStarRecovery}  for analyzing LSST's potential in recovering the light curves of pulsating variables and for defining the priority list of scientific cases to address from the very first LSST data releases.

\begin{acknowledgments}
This work was supported by: “Preparing for Astrophysics with the LSST Program" funded by the Heising-Simons Foundation, and administered by Las Cumbres Observatory with a grant for the publication and with the Kickstarter grant “Period and shape recovery of light curves of pulsating stars in different Galactic environments (KSI-8)”; Mini grant INAF 2022 “Are the Ultra Long Period Cepheids cosmological standard candles?" (PI: Musella, I.); Mini grant INAF 2022 “MOVIE@Rubin-LSST: enabling early science" (PI: Di Criscienzo, M.); INAF-ASTROFIT fellowship; Project PRIN MUR 2022 (code 2022ARWP9C) “Early Formation and Evolution of Bulge and HalO (EFEBHO)" (PI: Marconi, M.),  funded by European Union – Next Generation EU; Large grant INAF 2023 MOVIE (PI: M. Marconi) and
ASI-Gaia (“Missione Gaia Partecipazione italiana al DPAC – Operazioni e Attività di Analisi dati”); International Space Science Institute (ISSI) in Bern, through ISSI International Team project SHoT: The Stellar Path to the Ho Tension in the Gaia, TESS, LSST and JWST Era.
\end{acknowledgments}

\bibliography{main}{}
\bibliographystyle{aasjournal}

\end{document}